\documentclass{jpp}

\usepackage{amsmath}
\usepackage[utf8]{inputenc}
\usepackage{graphicx}
\usepackage{hyperref}
\usepackage{color}
\usepackage{booktabs}
\usepackage{siunitx}
\usepackage{overpic}

\title{Runaway electron generation during tokamak start-up}
\shorttitle{Runaways during tokamak start-up}
\author{M. Hoppe\aff{1,2}\corresp{\email{mathias.hoppe@epfl.ch}},  I. Ekmark\aff{1}, E. Berger\aff{1} \and T. Fülöp\aff{1}}
\shortauthor{M. Hoppe, I. Ekmark, E. Berger, and T. Fülöp}

\affiliation{
	\aff{1}Department of Physics, Chalmers University of Technology, SE-41296 Göteborg, Sweden
	\aff{2}Ecole Polytechnique Fédérale de Lausanne (EPFL), Swiss Plasma Center (SPC), CH-1015 Lausanne, Switzerland
}

\newcommand{\BKD}{\textsc{Bkd0}}
\newcommand{\DREAM}{\textsc{Dream}}
\newcommand{\DYON}{\textsc{Dyon}}
\newcommand{\LUKE}{\textsc{Luke}}
\newcommand{\SCENPLINT}{\textsc{Scenplint}}
\newcommand{\STREAM}{\textsc{Stream}}

\newcommand{\dd}{\mathrm{d}}
\newcommand{\ee}{\mathrm{e}}
\newcommand{\Eceff}{E_{\rm c}^{\rm eff}}
\newcommand{\Ec}{E_{\rm c}}
\newcommand{\ED}{E_{\rm D}}
\newcommand{\fu}{\alpha_\tau}
\newcommand{\gmax}{\gamma_{\rm max}}
\newcommand{\Ip}{I_{\rm p}}
\newcommand{\Ire}{I_{\rm re}}
\newcommand{\Iref}{I_{\rm ref}}
\newcommand{\Iw}{I_{\rm wall}}
\newcommand{\johm}{j_{\Omega}}
\newcommand{\jre}{j_{\rm re}}
\newcommand{\jtot}{j_{\rm tot}}
\newcommand{\Lp}{L_{\rm p}}
\newcommand{\Lw}{L_{\rm wall}}
\newcommand{\me}{m_{\rm e}}
\newcommand{\nel}{n_{\rm e}}
\newcommand{\nre}{n_{\rm re}}
\newcommand{\Rw}{R_{\rm wall}}
\newcommand{\Te}{T_{\rm e}}
\newcommand{\Ve}{V_{\rm loop,ext}}
\newcommand{\Vw}{V_{\rm loop,wall}}

\newcommand{\taure}{\tau_{\rm re}}
\newcommand{\taurepar}{\tau_{\rm re}^{\parallel}}
\newcommand{\tauredrf}{\tau_{\rm re}^{\rm drifts}}

\begin{document}
    \maketitle

    \begin{abstract}
		Tokamak start-up is characterized by  low electron densities and
strong electric fields, in order to quickly raise the plasma current
and temperature, allowing the plasma to fully ionize and magnetic flux
surfaces to form.  Such conditions are ideal for the formation of
superthermal electrons, which may reduce the efficiency of ohmic
heating and prevent the formation of a healthy thermal fusion
plasma. This is of particular concern in ITER where engineering
limitations put restrictions on the allowable electric fields and
limit the prefill densities during start-up. In this study, we present
a new 0D burn-through simulation tool called \STREAM\ (STart-up
Runaway Electron Analysis Model), which self-consistently evolves the
plasma density, temperature and electric field, while accounting for
the generation and loss of relativistic runaway electrons. After
verifying the burn-through model, we investigate conditions under which runaway
electrons can form during tokamak start-up as well as their effects on
the plasma initiation. We find that Dreicer generation plays a crucial
role in determining whether a discharge becomes runaway-dominated or not,
and that a large number of runaway electrons could limit the ohmic heating
of the plasma, thus preventing successful burn-through or further ramp-up
of the plasma current. The runaway generation can be suppressed by raising
the density via gas fuelling, but only if done sufficiently early. Otherwise
a large runaway seed may have already been built up, which can avalanche
even at relatively low electric fields and high densities.

    \end{abstract}

    \section{Introduction}\label{sec:intro}
	Operation in future fusion devices will have to ensure a start-up
scenario that is reliable, reproducible and low-risk.  To achieve
that, a solid understanding of all aspects of plasma initiation is
needed. If the operating parameters are not chosen with care,
plasma initiation can sometimes lead to the formation of a beam of
superthermal electrons, sometimes referred to as runaway
electrons  \citep{Knoepfel1979}.

Runaway beam formation is usually a significant concern in tokamak
disruptions, when the temperature falls and induces a large
electric field that accelerates electrons
\citep{Breizman_2019}. Runaway acceleration was
frequently observed during plasma initiation in the early days of
tokamak operation, but has received less attention lately, as most
current tokamaks can tune the operational parameters so that
runaway electron discharges are avoided. However, it is not clear
how the parameters should be chosen for future reactor-scale
tokamaks, like ITER. To avoid the risk of failed start-up and
corresponding delays or possible damage to in-vessel components caused
by runaway electrons, careful investigation of the plasma initiation
is needed.

In future tokamaks, the electric field applied for ionization and
to ramp up the plasma current is limited due to engineering
constraints related to the superconducting magnetic coils. In
ITER, the maximum electric field available for plasma breakdown is $0.3\;\rm V/m$  \citep{Gribov_2007}. This is considerably
lower than the typical value for the electric field available in
current devices ($\sim 1\;\rm V/m$). Tokamak start-up using a weak
electric field requires a low prefill gas pressure
\citep{de_Vries_2019}.  Due to the low collisionality at low
prefill pressure electrons are more prone to accelerate to high energies.
Therefore, plasma initiation using low prefill pressure is often
associated with discharges dominated by runaway electrons.

Tokamak plasma initiation consists of the breakdown phase, the
burn-through phase and the subsequent ramp-up of current. Plasma
breakdown by electron avalanche is induced by an applied electric
field \citep{Lloyd_1991}. In the breakdown phase, losses are dominated by
transport along magnetic field lines. After plasma breakdown, the next
phase is the burn-through, when ionization continues, as long as the
heating (that can consist of both ohmic and auxiliary heating)
overcomes losses due to ionization and radiation from the fusion fuel
and impurities \citep{Kim2012}.

Burn-through allows the plasma to reach full ionization, high
temperature and low resistance. After this phase, the plasma current
can be ramped-up efficiently by an applied electric field. As the
plasma current increases closed flux-surfaces are formed. Burn-through
is successful if the plasma reaches high enough ionization for the
heating power provided to be greater than the power lost due to ionization,
radiation and transport. The runaway electron content strongly affects the
efficiency of the burn-through and the electron density
development \citep{de_Vries_2020}. Runaway electrons may prevent further
increase in temperature by impairing the ohmic coupling between the electric
field and plasma, and thereby hinder the ramp-up of the plasma current.

Plasma initiation is an inherently dynamic situation in which many
plasma parameters (temperature, density, current, electric field
etc.)~evolve simultaneously and depend on each other, and is often
described using a set of coupled, non-linear differential
equations. The situation is complicated by the fact that the presence
of runaway electrons affects the plasma conductivity and the
ionization rate of atoms, hence altering the evolution of the plasma
parameters. It is therefore essential that start-up modelling tools include
runaway electron physics. However, most of the previously available numerical
solvers for the tokamak start-up problem do not include the effect of energetic
electrons \citep{Kim2020}. The code currently used for ITER start-up
development---the \SCENPLINT\ code~\citep{scenplint,transmak}---does include
effects of runaway electrons, but it uses simplified models for the runaway
generation rates. 

In this paper we investigate the effect of runaway electrons on
plasma initiation. In Sec.~\ref{sec:theory} we present the theoretical
model underlying the simulation tool \STREAM\footnote{The source code
is available at \url{https://github.com/chalmersplasmatheory/STREAM}.}\ 
({\em STart-up Runaway Electron Analysis Model}) that includes the essential
physical processes necessary for investigating start-up scenarios.
\STREAM\ computes the runaway electron generation
self-consistently with the plasma density, temperature, and ion
charge-state evolution, as well as the electric field evolution, 
models for coupling to the conducting structures in the wall, and transport along magnetic field
lines.    In Sec.~\ref{sec:benchmark} we show that the burn-through model
in \STREAM\ is in excellent agreement with the predictions of
\DYON~\citep{Kim2012}, both for ITER and JET parameters, when runaway
electrons are not considered. In Sec.~\ref{sec:iter} we demonstrate the
effect of runaway electrons on ITER burn-through scenarios. Finally, in
Sec.~\ref{sec:conclusions} we summarize our conclusions.

    \section{Burn-through model}\label{sec:theory}
    During the burn-through and ramp-up phases of tokamak start-up the
applied electric field is responsible for driving an ohmic plasma current
and thereby heating the plasma. As the temperature evolves,
the resistivity changes, which impacts the plasma current.
The atoms in the plasma chamber are initially primarily neutrals, but as the
burn-through proceeds, they rapidly ionize and contribute to radiation
loss processes, significantly affecting the temperature
evolution. Superthermal electrons can be accelerated during many stages of
the start-up, and their generation depends sensitively on background
plasma parameters such as the electric field strength, electron
temperature and density.

Building on the self-consistent disruption simulation tool
\DREAM~\citep{Hoppe2021}, a new tool \STREAM\ has been developed. The new tool 
introduces some physics essential to tokamak start-up which is not included in
\DREAM, and implements a 0D plasma model similar to the \DYON\ 
code~\citep{Kim2012}. Specifically, \STREAM\ evolves the quantities listed in 
Table~\ref{tab:quantities}, including the temperatures of electrons and an 
arbitrary number of ion species, the thermal-, runaway electron and ion charge 
state number densities, as well as the parallel electric field and the plasma
current through the ohmic and runaway electron current densities. In the
following section we describe the main elements of the model implemented in
\STREAM.

\begin{table}
    \centering
    \caption{Background plasma quantities evolved by \STREAM.} 
	\label{tab:quantities}
    \begin{tabular}{l l}
        \toprule
        Quantity & Description \\
        \midrule
            $E_\parallel$ & Parallel electric field in plasma\\
            $I_{\rm wall}$ & Current in conducting structures surrounding the plasma\\
            $j_{\Omega}$ & Ohmic current density\\
            $j_{\rm re}$ & Runaway electron current density\\
            $n_{\rm e}$ & Electron density\\
            $n_i^{(j)}$ & Density of ion species $i$, charge state $j$ ($j=0$ is neutral)\\
            $n_{\rm re}$ & Runaway electron density\\
            $T_{\rm e}$ & Electron temperature\\
			$T_i$ & Ion temperature\\
            $V_{\rm loop,wall}$ & Voltage in conducting structures surrounding the plasma\\
        \bottomrule
    \end{tabular}
\end{table}
\subsection{Ions and neutrals}
\begin{figure}
	\centering
	\includegraphics[width=0.5\textwidth]{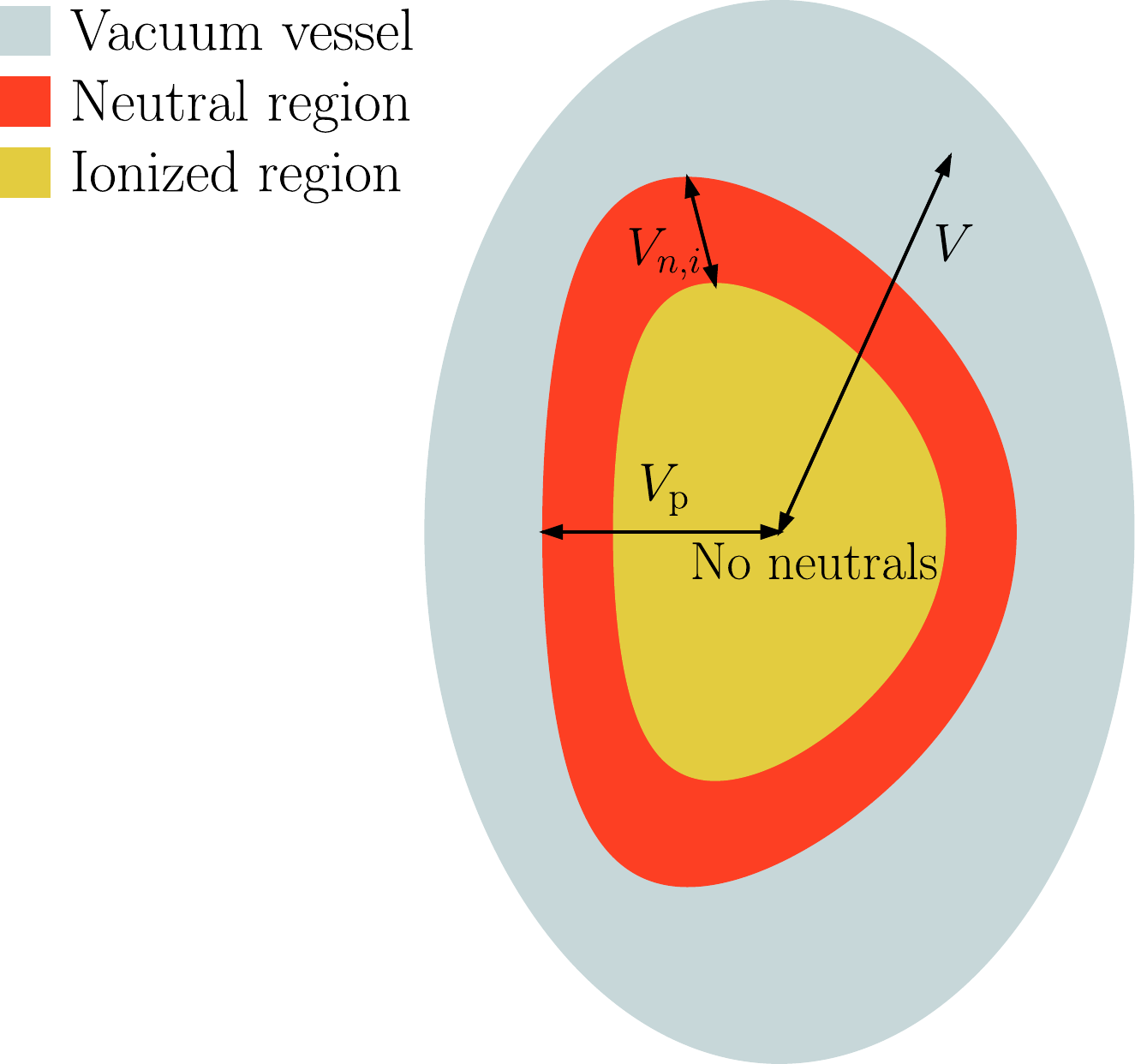}
	\caption{
        The two-volume model assumes that neutral particles move freely inside the
        vacuum vessel of volume $V$, except in the centre of the plasma
        volume $V_{\rm p}$ where plasma formation has come sufficiently far for
        neutrals to be fully screened out. The region of the plasma where
        neutrals and ions of species $i$ coexist has volume $V_{n,i}$.
	}
	\label{fig:twovolumes}
\end{figure}

\STREAM\ allows for an arbitrary number of ion species to be considered in the
simulation and evolves the density of each charge state separately. The fact
that the ionization mean-free path of neutrals decreases as the plasma develops,
so that neutrals can be screened out of the plasma core, is accounted for by using
the two-volume model introduced by \cite{Lloyd1996}. In the two-volume model,
illustrated in figure~\ref{fig:twovolumes}, the plasma is assumed to occupy a
volume $V_{\rm p}$, with neutrals of species $i$ only being able to penetrate a
sub-volume $V_{n,i}$ of the plasma. In addition, neutrals are assumed to be
homogeneously spread out in the volume consisting of $V_{n,i}$ and region
outside of the plasma. This means that the total volume occupied by the neutrals
of species $i$ is $\gamma_{n,i} V$, where the neutral volume coefficient is
given by 
\begin{equation}
    \gamma_{n,i} = 1 -\frac{V_{\rm p} - V_{n,i}}{V}.
\end{equation}
It is only the neutrals within the plasma which contribute to atomic processes.

The geometry of the plasma is specified using an analytic magnetic field model
described in appendix A of \cite{Hoppe2021}. The model includes the effects of
elongation and triangularity on the plasma. From the shaping
profiles prescribed, the plasma volume is computed numerically by
\begin{equation}
    V_{\rm p} = \int_0^a \dd r \int_0^{2 \pi} \dd\phi \int_{-\pi}^{\pi}  \mathcal{J}\dd\theta,
\end{equation}
where $a$ is the (possibly time-evolving) plasma minor radius and 
\begin{equation}
	\mathcal{J} = \frac{1}{|\nabla \phi \cdot (\nabla \theta \times \nabla r )|}
\end{equation}
is the spatial Jacobian, $r$ is the minor radius coordinate and $\phi$ and
$\theta$ are the toroidal and poloidal angles, respectively.

The volume occupied by the neutrals of species $i$ is determined from a formula
accounting for the plasma elongation $\kappa$ and triangularity $\delta$ by
subtracting the ionization mean-free path for species $i$ from the plasma
radius, so that
\begin{equation}
    V_{n,i} = 2\pi^2 R_0\kappa\left[ a^2 - \left(a-\lambda_i\right)^2 \right]
        + 2\kappa\delta\left(8-3\pi^2\right)\left[ a^3 - \frac{\left(a-\lambda_i\right)^3}{3} \right].
\end{equation}
The ionization mean-free path for neutrals of species $i$ is determined by
\citep{Lloyd1996}
\begin{equation}
    \lambda_i = \frac{v_{{\rm th},i}}{\nel I_i^{(0)}},
\end{equation}
with $v_{{\rm th},i} = \sqrt{2T_i/m_i}$ the thermal ion speed, $T_i$ and $m_i$
the temperature and mass respectively for species $i$, $\nel$ the 
electron density and $I_i^{(0)}$ the rate at which neutrals of species $i$ are
ionized. The ionization rate $I_i^{(0)}$ is taken from the Atomic Data and
Analysis Structure (ADAS) \citep{ADAS}.

\subsubsection{Particle balance}
The density of ions of species $i$ in charge state $j$ evolves through
ionization, recombination and charge exchange processes according to
\begin{equation}\label{eq:ions}
    \begin{aligned}
        \frac{\dd n_i^{(j)}}{\dd t} &= \frac{1}{V_i^{(j)}}\left[
            \hat{V}_i^{(j-1)} I_i^{(j-1)} \nel  n_i^{(j-1)} -
            \hat{V}_i^{(j)}I_i^{(j)} \nel n_i^{(j)}\right.\\
            &\left.+\hat{V}_i^{(j+1)} R_i^{(j+1)} \nel n_i^{(j+1)} -
            \hat{V}_i^{(j)} R_i^{(j)} \nel  n_i^{(j)} +
            \hat{V}_\star^{(0)} n_\star^{(0)} A_{i,{\rm cx}}^{(j)}
        \right] + S_i^{(j)}.
    \end{aligned}
\end{equation}
Here, $I_i^{(j)}$ is the rate of ionization of species $i$ from charge state
$j$ to $j+1$ and $R_i^{(j)}$ the rate of recombination of species $i$ from
charge state $j$ to $j-1$\footnote{With this notation,
$I_i^{(Z_i)}=R_i^{(0)}=0$.}, both of which are taken from ADAS. The total volume
$V_i^{(j)}$ occupied by ions of species $i$ in charge state $j$ is given by
\begin{equation}
    V_i^{(j)} = \begin{cases}
        \gamma_{n,i} V, &\quad j=0,\\
        V_{\rm p}, &\quad j \geq 1,
    \end{cases}
\end{equation}
while the volume {\em inside the plasma} $\hat{V}_i^{(j)}$ occupied by ions of
species $i$ and charge state $j$ is
\begin{equation}
    \hat{V}_i^{(j)} = \begin{cases}
        V_{n,i}, &\quad j=0,\\
        V_{\rm p}, &\quad j \geq 1.
    \end{cases}
\end{equation}
The interaction term $A_{i,{\rm cx}}^{(j)}$, for charge exchange
with neutrals of the main ion species (denoted with a star) and an
impurity ion, is
\begin{equation}
    A_{i,{\rm cx}}^{(j)} = \begin{cases}
        \left(-1\right)^{(j+1)} \sum_{k,l\geq 1} R_{ik,{\rm cx}}^{(l)} n_k^{(l)}, &\quad i=\star, k\neq \star,\\
        R_{ik,{\rm cx}}^{(j+1)} n_i^{(j+1)} - R_{ik,{\rm cx}}^{(j)}n_i^{(j)}, &\quad i\neq \star, k=\star,
    \end{cases}
\end{equation}
with $R_{ik,{\rm cx}}^{(j)}$ denoting the charge exchange rate, which is also
taken from ADAS.

The external particle source/sink $S_i^{(j)}$ models the influx of neutrals from
the wall, as well as the outflux of ions due to transport:
\begin{equation}
    S_i^{(j)} = \begin{cases}
        \Gamma_{i,{\rm in}}^{(0)}/V_i^{(0)}, &\quad j=0,\\
        -n_i^{(j)}/\tau_i, &\quad j\geq 1,
    \end{cases}
\end{equation}
where $\tau_i$ is the confinement time for species $i$. The neutral influx
$\Gamma_{i,{\rm in}}^{(0)}$ is generally given by
\begin{equation}\label{eq:neutral:influx}
    \Gamma_{i,{\rm in}}^{(0)} =
        V_{\rm p}\sum_k\sum_{l\geq 1}
        \frac{Y_k^i n_k^{(l)}}{\tau_k}.
\end{equation}
The sputter yield $Y_k^i$ for species $i$ due to the bombardment of
incident species $k$ is prescribed by the user.  The value of $Y_k^i$
depends on the underlying physical mechanism. The sputtering can be due to
chemical sputtering (e.g.\ in the case of carbon wall \citep{Kim2012}),
in which case $Y_k^i$ is approximately constant with a value based on
plasma surface experiment data, or physical sputtering, when $Y_k^i$ depends
on the incident ion energy (i.e.\ ITER wall) \citep{Kim_2013}.

The number of free electrons in the plasma is constrained via quasi-neutrality
to
\begin{equation}
    \nel  = \sum_{ij} Z_{0,i}^{(j)}  n_i^{(j)},
\end{equation}
where $Z_{0,i}^{(j)}$ denotes the net charge number of ion species $i$ in charge
state $j$, i.e. $Z_{0,i}^{(j)}=j$.

\subsubsection{Particle confinement}
The confinement time varies greatly during start-up and is determined by different
mechanisms during different phases of the start-up. In the early stages, before
closed flux surfaces have formed, the confinement is set by transport along
magnetic field lines to the wall. With an effective connection length
$L_{\rm f}$, the thermal particle confinement time can be estimated as
\citep{Kim2012}
\begin{equation}\label{eq:taupar}
    \tau_{i,\parallel} = \frac{L_{\rm f}}{C_{\rm s}},
\end{equation}
where $C_{\rm s} = \sqrt{(\Te+T_{\rm \star})/m_\star}$ is the main ion sound speed.
It was shown by \cite{Kim2012} that the same confinement time can also be used
for impurities when they are trace.

At low plasma currents, the effective connection length will depend on the
magnitude of stray magnetic fields. As the plasma current increases and
gradually exceeds the eddy currents in surrounding conducting structures, the
connection length increases and approaches infinity as closed flux surfaces
form. We therefore model the effective connection length as in \DYON\ 
\citep{Mineev2014,Kim2020}:
\begin{equation}\label{eq:connlen}
    L_{\rm f} = \frac{3a}{4}\frac{B_\phi}{B_z}\exp\left(\frac{\Ip}{\Iref}\right),
\end{equation}
where $B_\phi$ is the toroidal magnetic field strength, $\Ip$ the total plasma 
current and $\Iref$ denotes the plasma current where closed flux surfaces form
and the effect of stray magnetic fields greatly reduces. The stray field $B_z$
is composed of the vertical magnetic field $B_{\rm v}$ and a component
$B_{\rm eddy}$ resulting from eddy currents in conducting structures surrounding
the plasma:
\begin{equation}
    B_{\rm eddy} = \frac{\mu_0}{2\pi l_{\rm wall}}\Iw,
\end{equation}
where $\mu_0$ is the permeability of free space and $l_{\rm wall}$ denotes the
distance between the centre of the plasma and the conducting structure.

During later stages of the discharge, the thermal particle confinement is
expected to be dominated by turbulent transport. We can therefore estimate the
confinement time during the later stages using the Bohm scaling
\begin{equation}\label{eq:tauperp}
    \tau_{i,\perp} = \frac{a^2}{2D_{\rm Bohm}},
\end{equation}
where the Bohm diffusion coefficient $D_{\rm Bohm} = \Te/(16eB_\phi)$.
To allow the confinement time to smoothly transition from
equation~\eqref{eq:taupar} to~\eqref{eq:tauperp} during the start-up, we let the
total confinement time $\tau_i$ satisfy
\begin{equation}\label{eq:tau}
    \frac{1}{\tau_i} = \frac{1}{\tau_{i,\parallel}} + \frac{1}{\tau_{i,\perp}}.
\end{equation}

\subsection{Heat}
The temperature is evolved separately for each plasma species. The bulk electron
temperature $\Te$ is related to the electron thermal energy density
$W_{\rm e} = 3\nel \Te/2$, and the thermal energy density is in turn evolved according
to
\begin{equation}\label{eq:We}
    \frac{\dd W_{\rm e}}{\dd t} = j_\Omega E_\parallel -
        \nel \sum_i\sum_{j=0}^{Z_i} \frac{\hat{V}_i^{(j)}}{V_i^{(j)}}
		n_i^{(j)} L_i^{(j)} - \sum_i Q_{{\rm e}i} - \frac{W_{\rm e}}{\tau_{\rm e}}.
\end{equation}
The first term represents the ohmic heating provided by the electric field, and
the second term represents energy losses via inelastic atomic processes with
\begin{equation}
    L_i^{(j)} = L_{\rm line} + L_{\rm free} +
        \Delta W_i^{(j)}\left( I_i^{(j)} - R_i^{(j)} \right),
\end{equation}
where $L_{\rm line}$ is the radiated power by line radiation, $L_{\rm free}$
by recombination radiation and bremsstrahlung, and the other terms represent
the change in potential energy due to excitation and recombination, with the
same rate coefficients as in equation~\eqref{eq:ions}. Note that $R_i^{(j)}$
and $L_{\rm free}$ vanish for $j=0$, while $I_i^{(j)}$ vanishes for $j=Z_i$.
The ionization threshold $\Delta W_i^{(j)}$ is retrieved from the NIST
database \citep{NIST} while the other rate coefficients are taken from ADAS. The
collisional heat transfer $Q_{kl}$, which appears in the term for heat exchange
with ions in equation~\eqref{eq:We}, is generally given for two arbitrary plasma
species $k$ and $l$ by
\begin{equation}\label{eq:Qequil}
    Q_{kl} =
    \frac{\left\langle nZ^2 \right\rangle_k \left\langle nZ^2\right\rangle_l e^4\ln\Lambda_{kl}}
        {\left(2\pi\right)^{3/2}\epsilon_0^2 m_km_l}
    \frac{T_k - T_l}{\left(\frac{T_k}{m_k}+\frac{T_l}{m_l}\right)^{3/2}},
\end{equation}
with the weighted charge
$\langle nZ^2\rangle_i=\sum_{j=0}^{Z_i}n_i^{(j)}\left(Z_{0,i}^{(j)}\right)^2$.
Finally, the electron confinement time $\tau_{\rm e}$ is taken to be the same as that
of the main ions, given in equation~\eqref{eq:tau}, due to the requirement of
ambipolarity. The Coulomb logarithms are modelled using
\citep{Wesson2011}
\begin{subequations}
\begin{align}
	\ln\Lambda_{\rm ee} = \ln\Lambda_{{\rm e}i} &= 14.9 + \ln\left(\frac{\Te}{1\,\text{keV}}\right) - 0.5\ln\left(\frac{\nel}{10^{20}\,\text{m}^{-3}}\right),\\
	\ln\Lambda_{ii} &= 17.3 + \frac{3}{2}\ln\left(\frac{\Te}{1\,\text{keV}}\right) - 0.5\ln\left(\frac{\nel}{10^{20}\,\text{m}^{-3}}\right),
\end{align}
\end{subequations}
for electron-ion and ion-ion collisions, respectively. The thermal energy
density for each ion species evolves according to
\begin{equation}
	\begin{aligned}
		\frac{\dd W_i}{\dd t} &=
			\sum_k Q_{ik} -
			\frac{3}{2}\frac{\hat{V}_{\star}^{(0)}}{V_{\rm p}}
			n_\star^{(0)}\left( T_i - T_0 \right) R_{i,{\rm cx}}^{(1)}n_i^{(1)} -
			\frac{W_i}{\tau_i},
	\end{aligned}
\end{equation}
where $T_0 = \SI{300}{K}$ is the temperature of the main species neutrals,
$\tau_i$ is given by equation~\eqref{eq:tau}, and the sum in the first term runs
over all particle species, including electrons.

\subsection{Electric field}\label{sec:efield}
The toroidal electric field inside the plasma depends on the externally applied
loop voltage as well as the time rate of change of the plasma current. If the
plasma is surrounded by conducting structures, such as a metallic wall, the
plasma will be inductively coupled to these structures and the electric field
and current dynamics will be correspondingly affected. In \STREAM, we use the
same model for the electric field and current dynamics as in \DYON\ 
\citep{Kim2012}, described by one circuit equation for the plasma and one for
the conducting structure surrounding it:
\begin{subequations}
\label{eq:Vloop}
\begin{align}
	2\pi R_0 E_\parallel + \Lp\frac{\dd\Ip}{\dd t} + M\frac{\dd\Iw}{\dd t} &= \Ve,\\
	\Vw + \Lw\frac{\dd\Iw}{\dd t} + M\frac{\dd\Ip}{\dd t} &= \Ve,
\end{align}
\end{subequations}
where $\Ve$ is the externally applied loop voltage, $R_0$ is the tokamak
major radius, $E_\parallel$ the parallel electric field inside the plasma, $\Ip$
the total plasma current, $\Iw$ the current in the conducting structure and 
$\Vw=\Rw\Iw$ is the loop voltage in the conducting structure. The
inductances $\Lp$ (plasma inductance), $\Lw$ (wall inductance) and $M$
(plasma-wall mutual inductance), and the wall resistance $\Rw$, are free
parameters in the model.

Due to \STREAM's heritage from \DREAM, the total
plasma current $\Ip$ is obtained from the total current density $\jtot$ in
the system by multiplying with the plasma cross-sectional area.
An important difference between \STREAM\ and \DYON\ is that in \STREAM\ the current
density in turn consists of an ohmic component $\johm = \sigma E_\parallel$,
where $\sigma$ is the conductivity of the plasma as determined
by~\cite{Redl2021}, and a runaway electron component $\jre = ec\nre$, with $e$
the elementary charge, $c$ the speed of light, and $\nre$ the runaway electron
density. The calculation of the runaway electron density $\nre$ is described in
detail in section~\ref{sec:runaways}.

\subsection{Runaway electrons}\label{sec:runaways}
Electrons are said to ``run away'' when the collisional friction acting on them
is weaker than other accelerating forces. In most cases, the accelerating force
is an electric field, and in order for it to provide net acceleration to
electrons it must exceed the Connor-Hastie threshold \citep{Connor_1975}
\begin{equation}
	E > E_{\rm c} = \frac{e^3\nel\ln\Lambda_{\rm ee}}{4\pi\epsilon_0^2\me c^2},
\end{equation}
where $\epsilon_0$ is the vacuum permittivity and $\me$ the electron rest mass.
Above this threshold, all electrons with relativistic momentum
$p > p_{\rm c}\approx1/\sqrt{E/E_{\rm c}-1}$ will be freely accelerated and run
away. Electrons can find themselves above $p=p_{\rm c}$, in the so-called
runaway region, in a number of ways. For example, they can enter this region in
momentum space through the collisionally diffusive leak from the thermal
population at a steady rate---the so-called Dreicer generation 
\citep{Dreicer1959}, that is exponentially sensitive to the electric field
normalized to the Dreicer field
\begin{equation}\label{Dreicer}
	\ED= \Ec\frac{\me c^2}{\Te} = \frac{e^3\nel\ln\Lambda_{\rm ee}}{4 \pi \epsilon_0^2 \Te}.
\end{equation}
When $E\approx 0.215\ED$, the electric field acceleration exceeds the
collisional friction for all electrons, leading to slide-away and a distribution
that is far from thermal. Slide-away can also occur at lower values of $E$ since
the Dreicer generation mechanism will gradually drain the thermal bulk of
particles and pull the electrons into the runaway region.

Existing runaway electrons can also create new ones through close
collisions with thermal electrons \citep{jayakumar}. This leads to an
exponential growth of the number of runaway electrons---an
avalanche. The avalanche mechanism has caused much concern for the potentially
highly aggressive growth of the number of runaway electrons during tokamak
disruptions \citep{Boozer2015,Breizman_2019}, which is seen as almost
inevitable in future reactors due to the presence of additional primary sources
of runaway electrons. The first of these, labelled the ``hot-tail'' mechanism,
occurs during the thermal quench of a disruption as the fastest electrons of the
original hot thermal distribution take longer to slow down. They may therefore
be accelerated by the electric field which is induced as the plasma
current drops due to the increased resistivity, before they have time to
thermalize.
During tritium
operation, energetic electrons resulting from tritium decay and
Compton scattering of $\gamma$ photons originating from the activated
wall are predicted to also provide a significant seed electron
population to drive the avalanche multiplication
\citep{MartinSolis2017,elongation_2020,vallhagen_2020}.

During tokamak start-up, the Dreicer and avalanche mechanisms are
expected to dominate the generation of runaway electrons, while the
other generation mechanisms do not contribute significantly. Hot-tail
generation is only expected to occur in cooling plasmas, and is not
relevant to the conditions of the initiating plasma. Tritium decay and
Compton scattering sources of runaway electrons are typically much
lower than Dreicer generation due to the large normalized electric
field $E/\ED$ and low electron densities that are typical during plasma
initiation.

\subsubsection{Runaway electron model}\label{sec:remodel}
Runaway electrons primarily influence the plasma evolution during start-up by
contributing a relativistic current component to the plasma current. Since the
distribution of runaways is usually strongly beamed along the magnetic field
lines, with speeds close to the speed of light, the runaways contribute a
current density $\jre=ec\nre$. This current density will in turn affect the
evolution of the loop voltage in equation~\eqref{eq:Vloop} and, by extension,
the ohmic current and heating.

The runaway electron density is evolved according to
\begin{equation}\label{eq:dnredt}
    \frac{\partial\nre}{\partial t} = \gamma_{\rm Dreicer} + \Gamma_{\rm ava}\nre - \frac{\nre}{\tau_{\rm re}},
\end{equation}
where $\gamma_{\rm Dreicer}$ is the rate at which electrons are generated by the
Dreicer mechanism~\citep{Dreicer1959}, $\Gamma_{\rm ava}$ is the
avalanche growth rate by which runaway electrons exponentially
multiply~\citep{Rosenbluth1997}, and $\tau_{\rm re}$ is the
runaway electron confinement time. The Dreicer generation rate is evaluated
using the neural network developed by~\cite{Hesslow2019nn}, while the avalanche
growth rate is given by the semi-analytical formula~\citep{Hesslow2019ava}
\begin{equation}\label{eq:GammaAva}
    \begin{gathered}
        \Gamma_{\rm ava} = \frac{e}{\me c\ln\Lambda_{\rm c}}
            \frac{\nel^{\rm tot}}{\nel}
            \frac{E_\parallel-\Eceff}
                {\sqrt{4+\bar{\nu}_{\rm s}\left(p_\star\right)\bar{\nu}_{\rm D}\left(p_\star\right)}},\\
        p_\star =
            \frac{\sqrt[4]{\bar{\nu}_{\rm s}\left(p_\star\right)\bar{\nu}_{\rm D}\left(p_\star\right)}}
                {\sqrt{E_\parallel/E_{\rm c}}},\\
		\ln\Lambda_{\rm c} = 14.6 + 0.5\ln\left(
			\frac{\Te\,[{\rm keV}]}
			{\nel\,[10^{20}\,{\rm m}^{-3}]}
		\right),
    \end{gathered}
\end{equation}
where $\nel^{\rm tot}$ is the total density of electrons (bound and
free), $\ln\Lambda_{\rm c}$ is a generalized Coulomb logarithm evaluated at
relativistic energies, and $\bar{\nu}_{\rm s}$ and $\bar{\nu}_{\rm D}$ are the
normalized slowing-down and deflection frequencies as defined
by~\cite{Hesslow2018}. The effective critical electric field $\Eceff$
is calculated as described in appendix C.2 of~\citep{Hoppe2021}, and
takes into account the effect of bremsstrahlung and synchrotron
radiation based on the approach of \cite{Hesslow_2018}. In partially
ionized plasmas, both the Dreicer generation rate and the avalanche
growth rate depend on the extent to which fast electrons can penetrate
the bound electron cloud around the impurity ion. This effect of
partial screening is taken into account in both the neural network for
Dreicer generation and in equation~\eqref{eq:GammaAva} for the
avalanche growth rate.

Using state-of-the-art models for the effect of partial screening is the main
difference between the runaway generation models used in the \SCENPLINT\ code
and \STREAM. As shown by \cite{Hesslow2019nn} these effects can lead to orders
of magnitude differences in the Dreicer generation rate, in plasmas containing
partially ionized atoms. Also, the generalized expression for the avalanche
growth rate has been shown to lead to large differences in the final runaway
current, compared to previously used expressions (e.g.~in
\cite{MartinSolis2017}), as demonstrated in
\cite{Hesslow2019ava,vallhagen_2020}.

As for ions, the runaway electron confinement time varies significantly
during start-up~\citep{Kavin2017}. In the early stages, before flux surfaces
have formed, the transport will be dominated by parallel transport, and the
distance traversed by a runaway before leaving the plasma is the same as for the
thermal particles, i.e.\ the connection length given in
equation~\eqref{eq:connlen}.
Starting from the equation of motion for the electron,
\begin{equation}\label{eq:mo}
    	\frac{\dd p}{\dd t} = eE_\parallel,
\end{equation}
where $p$ is the electron momentum, and letting $E_\parallel$ be positive in the
co-current direction, we can estimate its confinement time. By assuming $E$ to
be constant in time, integrating equation~\eqref{eq:mo} and solving for the
electron speed $v$, we obtain
\begin{equation}
	v \equiv \frac{\dd s}{\dd t} = \frac{eE_\parallel t/\me}{\sqrt{1+\left(\frac{eE_\parallel t}{\me c}\right)^2}}.
\end{equation}
This can in turn be integrated over the time period $\taurepar$ it takes for the
electron to travel a distance $L_{\rm f}$, to obtain the relation
\begin{equation}
	L_{\rm f} = \frac{\me c^2}{eE_\parallel}\left[
		\sqrt{1+\left(\frac{eE_\parallel\taurepar}{\me c}\right)^2} - 1
	\right],
\end{equation}
which can be solved for the runaway electron confinement time
\begin{equation}\label{eq:tau1}
	\taurepar = \frac{\me c}{eE_\parallel}\sqrt{\left(\frac{eE_\parallel L_{\rm f}}{\me c^2}+1\right)^2-1}.
\end{equation}

After flux surfaces have formed, the confinement time is rather set by the rate
at which the runaways gain energy and drift out of the plasma. Assuming that a
runaway electron drifts out of the plasma when reaching a relativistic energy
$\me c^2\gmax$, and assuming the electric field to be roughly constant during
acceleration, the confinement time in this second phase can be related to the
runaway electron energy via
\begin{equation}
    p_{\rm max} = \int_0^{\tauredrf} eE_\parallel\,\dd t
    = eE_\parallel\tauredrf.
\end{equation}
The maximum energy $\gmax=\sqrt{1+p_{\rm max}^2/\me^2c^2}$ for an electron was
estimated by~\cite{Knoepfel1979} as
\begin{equation}
    \gmax\approx\frac{56R_0}{a}I_{\rm p}\,[\mathrm{MA}],
\end{equation}
where $a$ and $R_0$ are the plasma minor and major radii respectively. The
runaway electron confinement time due to drifts can therefore be taken as
\begin{equation}\label{eq:tau2}
    \tauredrf\,[\mathrm{s}]\approx\frac{R_0}{10a}
    \frac{I_{\rm p}\,[\mathrm{MA}]}{E_\parallel\,[\mathrm{V/m}]}.
\end{equation}
Since the transition from the first to the second stage is determined by when
closed flux surfaces form, which depends on the development of the plasma
current, we interpolate between the confinement times~\eqref{eq:tau1}
and~\eqref{eq:tau2} using
\begin{equation}
    \frac{1}{\taure} =
        \frac{\exp\left(-I_{\rm p}/I_{\rm ref}\right)}{\taurepar} +
        \frac{1-\exp\left(-I_{\rm p}/I_{\rm ref}\right)}{\tauredrf},
\end{equation}
with $I_{\rm ref}$ the current at which flux surfaces form. $I_{\rm
ref}$ is assumed to be approximately  $\SI{100}{kA}$ in ITER \citep{Kim2020}.

    \section{Verification of the burn-through model}
	\label{sec:benchmark}The plasma initiation model presented in sec.~\ref{sec:theory} involves coupled
equations, most of which are non-trivial in their numerical
implementation. To verify that the implementation of \STREAM\ is
correct, we must compare its predictions to previously established
results.  The runaway electron generation models included in \STREAM\
are inherited from \DREAM, for which extensive benchmarks to previous
results have been made (some of which are documented
in \citep{Hoppe2021}).

For \STREAM, the crucial part to verify is the burn-through model,
excluding the runaway electron physics. Such models have been
implemented in several codes before, including in the
\DYON~\citep{Kim2012}, \SCENPLINT~\citep{scenplint,transmak} and
\BKD~\citep{Granucci} codes, which were carefully benchmarked against each
other recently in \citep{Kim2020}.  In this section we will reproduce the
results of the two burn-through scenarios considered in \citep{Kim2020}
with \STREAM. In the first scenario, a pure hydrogen plasma start-up
in an ITER-like setting is considered, assuming a perfectly insulating
vacuum vessel. In the second scenario, a more advanced JET case is
studied, involving multiple impurity species and a conducting
structure which affects the current and electric field evolution.

Although the burn-through models in \STREAM\ and \DYON\ are similar
regarding all the important physics components: the particle and heat
balance, the electrical circuit model, the neutral screening, and the
impurity charge-state evolution, there are slight differences.
While \DYON\ assumes all the ions to have the same temperature,
in \STREAM\ we allow for different ion species to have different
temperatures and only assume that the temperature is the same in every
charge state. Furthermore, there are also differences in how the
conductivity is calculated as described in sec.~\ref{sec:efield}. However, as
we shall see, in spite of these subtle differences, the agreement between
the results of the two codes is very good.

\subsection{ITER scenario}
We start by considering an idealized ITER ohmic burn-through
scenario with constant input parameters, similar to the one considered
in Sec.~2 of \cite{Kim2020}.  We assume a circular plasma and a
constant loop voltage corresponding to $\SI{0.3}{V/m}$.
The prefilled fuel gas pressure is $\SI{0.8}{mPa}$, corresponding
to an initial hydrogen density of $\SI{3.84e17}{m^{-3}}$. The
gas consists of pure hydrogen (no impurities) at
$T_i=\SI{0.026}{eV}$ (room temperature) in the beginning of the
simulation. The plasma inductance is calculated from $L=\mu_0
R_{\rm 0} [\ln{(8R_{\rm 0}/a)}+l_{\rm i}/2-2]$, where
$l_{\rm i}=0.5$ is assumed for the internal inductance and we neglect all currents in
surrounding passive structures. The wall recycling coefficient for hydrogen
$Y_{\rm H}^{\rm H}$ is set equal to one (no external gas fuelling). The input parameters are 
summarized in Table~\ref{tab:mergedparam}. Note that while hydrogen is used as
the main ion species, \STREAM\ has been temporarily modified in these
simulations to use charge-exchange rates for deuterium, which should correspond
more closely to the charge-exchange rates used for this scenario
in \citep{Kim2020}.

\begin{table}
    \centering
    \caption{\label{tab:mergedparam}
		Simulation parameters for the ITER and JET benchmark cases. Some
		parameters appear only in the models used in the JET case.
    }
    \begin{tabular}{l l l l}
        \toprule
        Parameter & Name & ITER & JET \\
        \midrule
            $p_\textrm{prefill}$  & Prefill gas pressure                  &  \SI{0.8}{mPa}  & \SI{2.7}{mPa}                      \\ 
            $\gamma           $   & Initial ionization degree             & \SI{0.2}{\percent}   & \SI{0.2}{\percent}                   \\
            $B_\textrm{tor}$      & Toroidal magnetic field               & \SI{2.65}{\tesla}  & \SI{2.4}{\tesla}                     \\
            $R_\textrm{0}$        & Plasma major radius                   & \SI{5.65}{\meter} & \SI{2.96}{\meter}                    \\
            $a$                   & Plasma minor radius                   & \SI{1.6}{\meter}  & See fig.~10a in \cite{Kim2020}                 \\             $d_\textrm{RS}$       & Passive structure-distance            & ---   & \SI{1}{\meter}                       \\ 
            $V_\textrm{vessel}$   & Vessel volume                         & \SI{1000}{\meter\cubed} & \SI{100}{\meter\cubed}               \\
            $T_\textrm{e}$        & Initial electron temperature           & \SI{1}{\electronvolt}  & \SI{1}{\electronvolt}                \\
            $T_i$        & Initial ion temperature               & \SI{0.026}{\electronvolt}& \SI{0.026}{\electronvolt}            \\
            $I_\textrm{p}$        & Initial plasma current                & \SI{2.4}{\kilo\ampere} & \SI{2.4}{\kilo\ampere}               \\
            $\Ve$                 & Loop voltage                          & \SI{12}{\volt}        & See fig.~10b in \cite{Kim2020}                \\ %
            $L_\textrm{p}$        & Plasma inductance                     &  \SI{11.3}{\micro\henry}    & \SI{5.19}{\micro\henry}              \\
            $L_\textrm{wall}$     & Wall inductance                       & --- & \SI{9.1}{\micro\henry}               \\
            $M$                   & Mutual inductance    &                  ---          & \SI{2.49}{\micro\henry}              \\
            $R_\textrm{wall}$     & Wall resistance                       & ---   & \SI{0.75}{\milli\ohm}                \\
            $c_1$                 & First recycling coefficient           & --- & 1.1                                  \\
            $c_2$                 & Second recycling coefficient          & --- & 0.05                                 \\
            $c_3$                 & Third recycling coefficient           & --- & 0.1                                  \\ 
            $f_\textrm{O}$        & Initial fraction of oxygen            & --- & 0.001                                \\ 
            $f_\textrm{C}$        & Initial fraction of carbon            & --- & 0                                    \\ 
        \bottomrule
    \end{tabular}
\end{table}

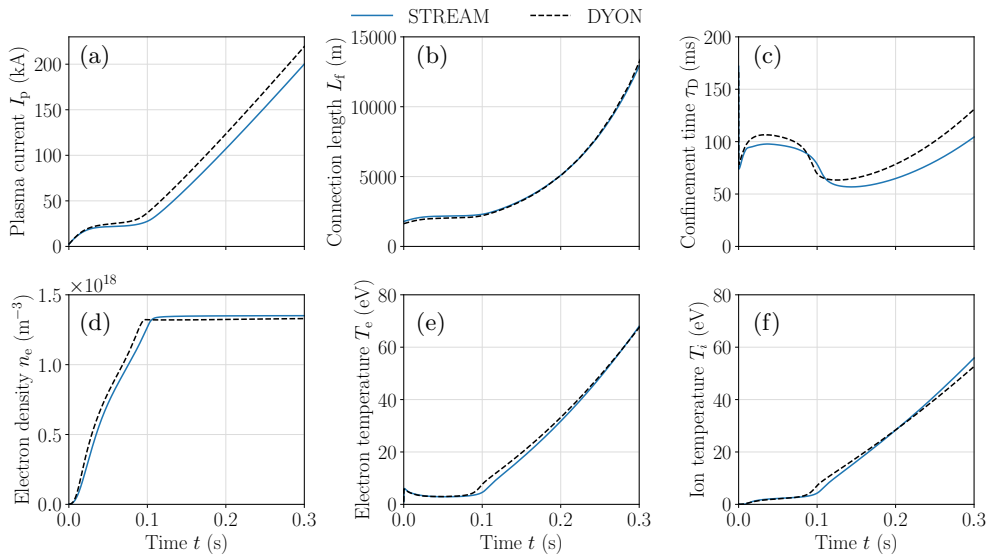
\begin{figure}
    \centering
    \begin{overpic}[width=\textwidth]{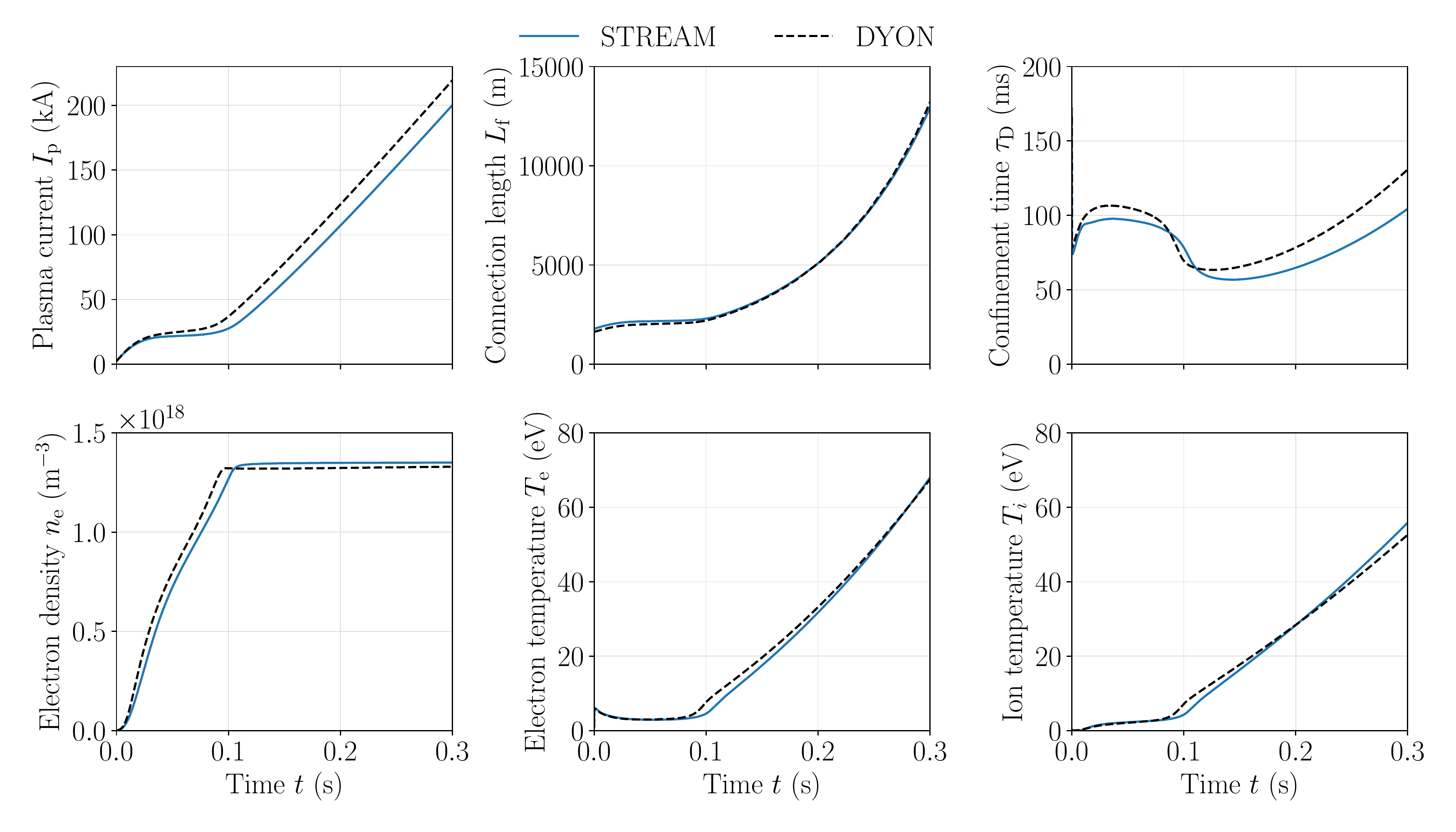}
		\put(9,50){(a)}
		\put(42,50){(b)}
		\put(75,50){(c)}
		\put(9,24){(d)}
		\put(42,24){(e)}
		\put(75,24){(f)}
	\end{overpic}
    \caption{
        Time evolution of plasma parameters for the idealized ITER scenario:
		(a) plasma current, (b) effective connection length, (c) confinement
		time, (d) electron density, (e) electron temperature, and (f) ion
		temperature. The input parameters are given in
		table~\ref{tab:mergedparam}. Solid blue line is obtained by \STREAM,
		dashed black line by \DYON.
    }
    \label{ITERKim}
\end{figure}

Figure \ref{ITERKim} shows the time-evolution of the plasma current, effective
connection length, particle confinement time, electron density, electron
temperature, and ion temperature. After modifying \STREAM\ to use the ADAS
charge-exchange rates for deuterium also for hydrogen, as was done in
\citep{Kim2020}, good agreement was found between the simulation results of
\DYON\ and \STREAM. Only a slightly delayed burn-through is seen in \STREAM\
which leads to a delayed rise of the plasma current, and a slightly faster
heating rate for both ions and electrons in \STREAM. These small differences
observed could potentially be explained by the different conductivities used
(the \STREAM\ simulations use the conductivity of \cite{Redl2021} while the
\DYON\ simulations use the Spitzer conductivity), the use of different line
radiation and charge exchange rates (the \STREAM\ simulations use ADAS rates
while the \DYON\ simulations use analytical fits), and differences in the
electron-ion equilibration term~\eqref{eq:Qequil} (the \DYON\ simulations use a
constant value $\ln\Lambda=10$ for the Coulomb logarithm).

\subsection{JET scenario}
Next, we consider an advanced burn-through scenario of a deuterium plasma with multiple
impurity species and a conducting structure near the plasma. As in
\citep{Kim2020}, we use parameters from a JET discharge with carbon wall, which
was used to validate the \DYON\ code. In this case, both the breakdown region
and the loop voltage are time-dependent, and the data is given in Figure 10 of
\citep{Kim2020}. The plasma minor radius is inferred from the plasma volume
given in Fig.~10a. The eddy current is calculated with the two-rings circuit
model in equation~\eqref{eq:Vloop}. To obtain agreement it was necessary to
modify the effective connection length according to $L_{\rm f}\to L_{\rm f}/3$
from equation~\eqref{eq:connlen} in \STREAM\ to use the exact same model as was
used in \citep{Kim2020}.

In the simulation, a time-evolving deuterium recycling coefficient is used.
The coefficient is given generally by
\begin{equation}
	Y_{\rm D}^{\rm D}(t) = c_1 - c_2\left(1 - \ee^{-t/c_3}\right),
\end{equation}
and in this scenario we set $c_1=1.1$, $c_2=0.05$ and $c_3=\SI{0.1}{s}$.

\begin{figure}
    \centering
    \includegraphics[width=\textwidth]{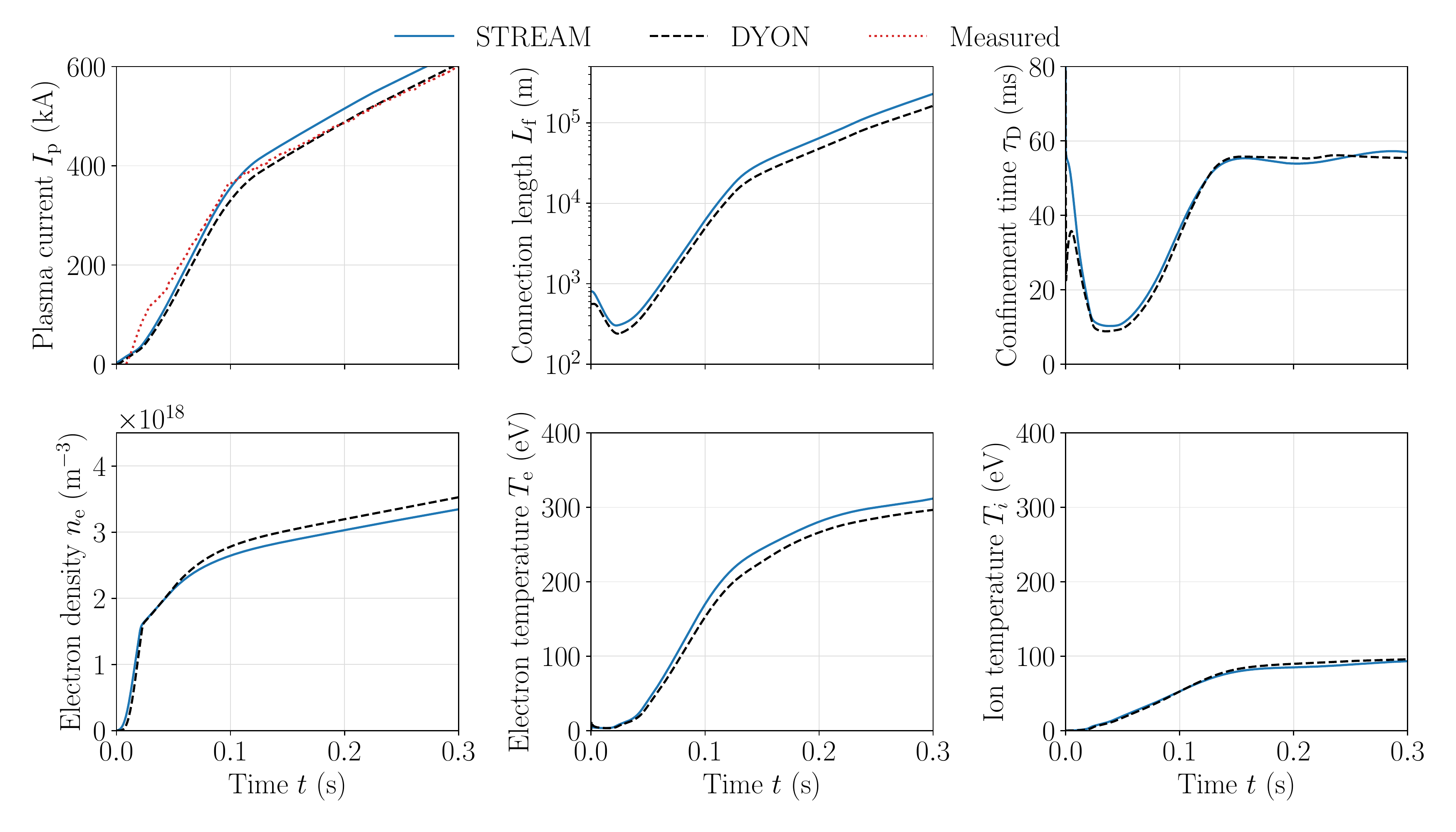}
    \caption{
		Time evolution of plasma parameters for the JET carbon wall discharge
		\#$77\,210$: (a) Plasma current, (b) electron density, (c) electron
		temperature, (d) effective connection length (e) ion temperature and (f)
		confinement time. The input parameters are given in
		Table~\ref{tab:mergedparam}. Solid blue line is obtained by \STREAM,
		dashed black line by \DYON, and dotted line is the measured plasma
		current.
	}
	\label{JETKim}
\end{figure}

Figure \ref{JETKim} shows the time-evolution of the plasma current, effective
connection length, particle confinement time, electron density, electron
temperature, and ion temperature. Also in this case \STREAM\ and \DYON\ results
agree well, with a slight deviation in the effective connection length
$L_{\rm f}$, and at early times in the deuterium confinement time
$\tau_{\rm D}$, possibly explaining differences between the two codes. The
differences in $L_{\rm f}$ and $\tau_{\rm D}$ could in turn possibly be
explained by differences in the evolution of the plasma volume and the plasma
major radius. The plasma volume was determined from figure 10a of
\citep{Kim2020}, while the plasma major radius was assumed constant. The plasma
volume plays an important role for the density and temperature evolution, while
the plasma major radius affects the value of the toroidal magnetic field used
in the simulation, all of which in turn affect the effective connection length
and the particle confinement time.

    \section{ITER burn-through simulations with runaway electrons}
    \label{sec:iter}The role of runaway electrons during tokamak start-up has 
been studied both experimentally~\citep{Esposito_1996,Yoshino_1997,Esposito_2003,de_Vries_2020} and theoretically \citep{Sharma_1988,de_Vries_2019}, and it is well known that the
presence of superthermal or runaway electrons during start-up can lead
to burn-through failing or the creation of a relativistic electron
beam.  Here, we use ``superthermal'' to mean electrons which are
moving at speeds much faster than the thermal speed, but not
undergoing free acceleration. By ``runaway electrons'', on the other
hand, we mean electrons which are sufficiently fast for the electric
field acceleration to dominate collisional slowing-down, and which
therefore accelerate freely. To study the dynamics of the former, one
generally needs to solve a kinetic equation which accounts for the
momentum dynamics of the electrons, while the latter can often be
studied using conceptually simpler fluid models (see e.g.~\cite{vallhagen_2020}). In this
paper, we will only consider runaway electron discharges and leave
considerations concerning superthermal discharges for a future
publication.

In this section we consider runaway electron generation in an ITER ohmic
first-plasma scenario. We investigate conditions for the appearance of runaway
electrons in the plasma as well as their effect on the plasma start-up.

\subsection{ITER first-plasma scenario}\label{sec:iterfirst}
We begin by considering runaway generation in an ITER ohmic first
plasma scenario, with the parameters given in the first column of
Table \ref{tab:mergedparam} and a deuterium-deuterium recycling
coefficient $Y_{\rm D}^{\rm D}=1$. Note, these parameters were chosen for benchmarking
purposes by \cite{Kim2020}, and they are quantitatively different from an actual
plasma initiation scenario in ITER. In this baseline scenario, the plasma 
volume, position, and loop voltage are held constant while in reality they
should be evolving in time, and no impurities are considered. Furthermore, the
vessel volume is assumed to be as in ITER with a fully completed first wall,
i.e.\ much smaller than the vessel volume in 2025 ITER first plasma.
Nevertheless, since we are primarily interested in qualitatively studying the
prevalence of runaway electrons in startup scenarios, we use these parameters to
reduce the complexity of the simulations.

\begin{figure}
    \centering
    \includegraphics[width=\textwidth]{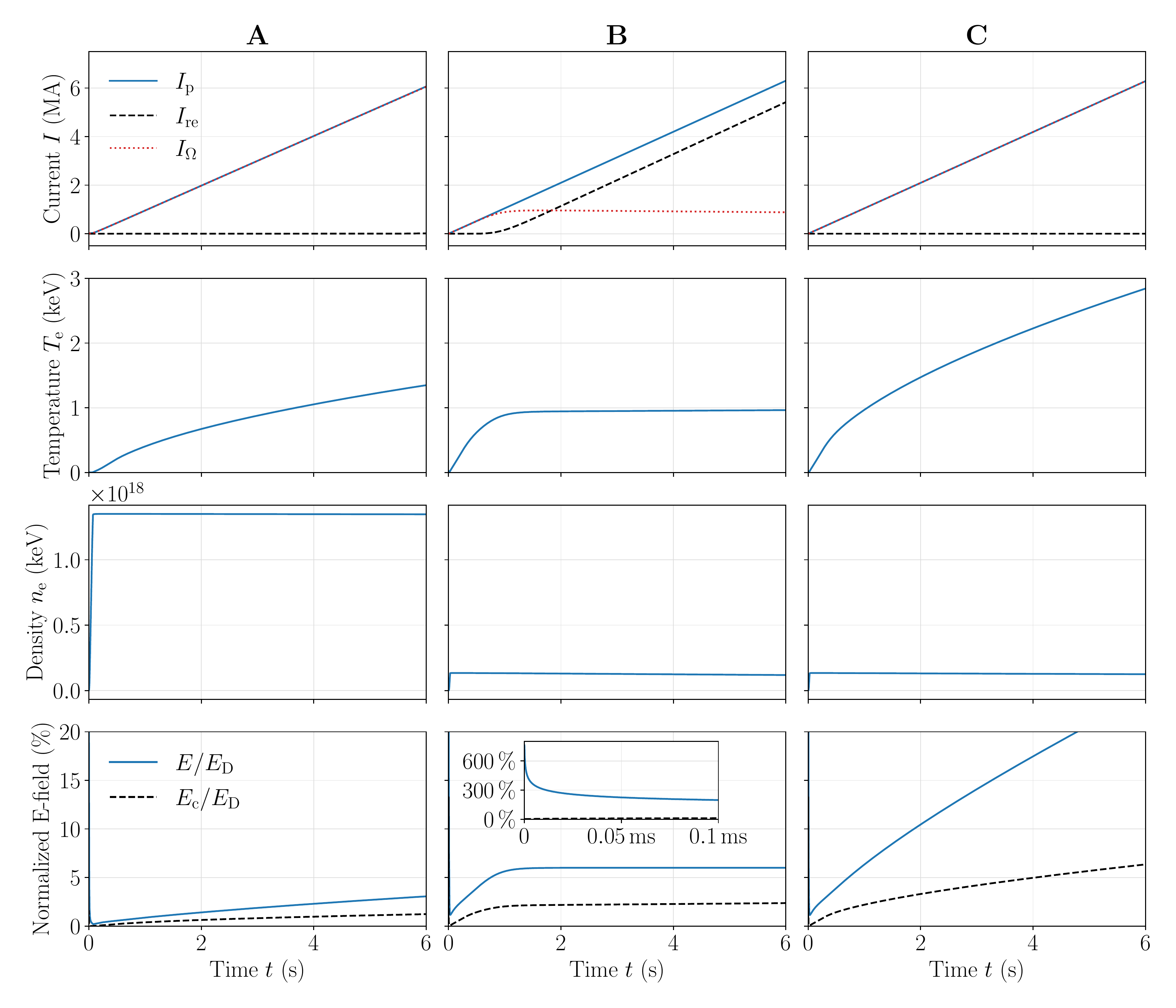}
    \caption{
        Time evolution of plasma parameters for the ITER first plasma scenario.
		Column A shows the baseline case, when no significant runaway current is
		generated. Column B shows a case when a significant runaway current is
		generated. Column C shows the same case as column B, but with runaway
		generation turned off. Here, $I_{\rm p}$ denotes the total plasma
		current, $I_{\rm re}$ the runaway electron current, and $I_{\Omega}$ the
		ohmic current.
	}
	\label{fig:REabc}
\end{figure}

We focus on the tokamak start-up from just after
breakdown when $\SI{0.2}{\percent}$ of deuterium atoms are ionised, the plasma
current has reached $\Ip=\SI{2.4}{kA}$ and the electron temperature is
$\Te=\SI{1}{eV}$. In this situation, closed flux surfaces have yet to
be formed and deuterium burn-through has not yet occurred. Column A in
Fig.~\ref{fig:REabc} illustrates the evolution of plasma current,
temperature and electric field.  Burn-through is
achieved within the first $\SI{100}{ms}$, after which the temperature
continues to increase and the plasma current is ramped up linearly.
In this case, no appreciable number of runaway electrons are
generated, despite the fact that $E/\ED$ takes values as high as
\SI{85}{\percent} early in the start-up, and remains well above the threshold
value $\Ec$ throughout the start-up phase. The reason for this is the strong
transport which expels all runaways that are generated before closed flux
surfaces have formed. By the time the flux surfaces are closed, the free
electron density and temperature have risen sufficiently for the ratio $E/\ED$
to be negligibly small. The subsequent increase in $E/\ED$ is due to the rising
temperature, but never reaches values large enough for significant Dreicer
generation to occur.

\begin{figure}
    \centering
    \begin{overpic}[width=0.67\textwidth]{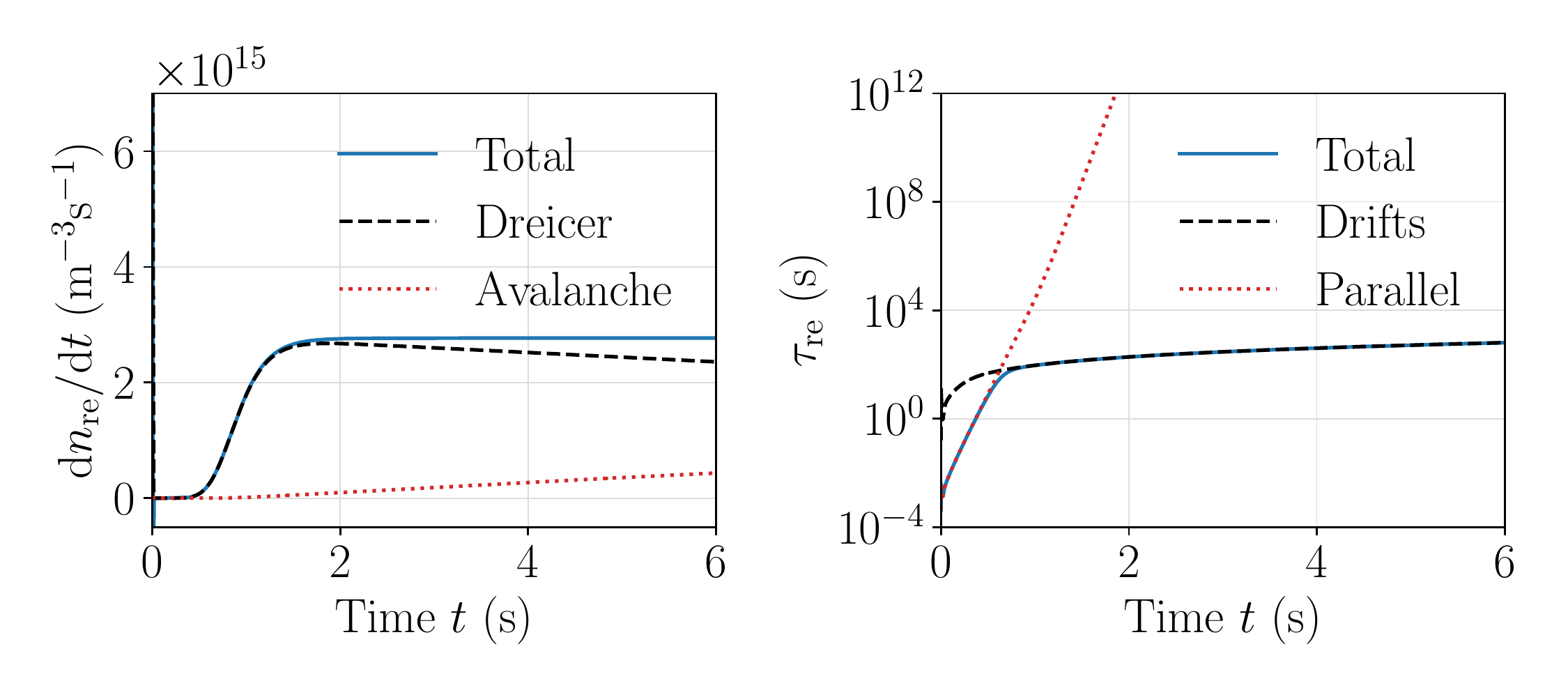}
		\put(11,35){(a)}
		\put(61,35){(b)}
	\end{overpic}
    \caption{
        Time evolution of runaway generation and runaway electron  confinement 
        time for case B.
	}
	\label{fig:REabc_confinementtime}
\end{figure}

By lowering the prefill pressure $p_{\rm prefill}$ by a factor of ten, to
$\SI{0.08}{mPa}$, so that the initial electron density is an order of magnitude
lower than in the baseline case, we obtain a scenario in which burn-through
succeeds and significant runaway electron generation occurs. In column B of
figure~\ref{fig:REabc} the evolution in this lower-prefill case is illustrated.
As in the baseline case, column A, the plasma current rises at the expected
rate, but the composition of the current is very different. The ohmic current
saturates at a value of about $\SI{1}{MA}$ while the current carried by the
runaway electrons continues to rise unhindered. As with the ohmic current, the
electron temperature also levels off since most of the energy transferred via
the electric field is deposited to the runaway electrons which, due to their low
rate of collisions, retain most of the energy without transmitting it to the
thermal electrons. This also causes $\ED$ to reach a steady value, and since the
externally applied loop voltage is held constant, the parameter $E/\ED$
saturates. It should be pointed out than in ITER, the loop voltage will
eventually be forced to decrease, thus potentially greatly reducing the runaway
generation. As long as $E>\Ec$, runaways can continue to reproduce through the
avalanche mechanism, requiring that the model be coupled to a simulation of the
current in surrounding coils for a quantitative study.

Figure~\ref{fig:REabc_confinementtime}a shows the role played by the Dreicer
and avalanche generation mechanisms, respectively, in the low-prefill pressure
case. Throughout the simulation, Dreicer generation remains the dominant
mechanism while avalanche only starts producing a noticeable number of runaways
after a few seconds, when a sufficient seed population has built up. The reason
for the strong Dreicer production is the large value of $E/\ED$ which is
sustained for a long time. This is in contrast to the typical situation in
disruptions where the electric field usually attains significant fractions of
$\ED$ only for a brief period during the current quench, after which it
quickly drops and approaches $\Eceff$, driving further runaway production
mainly via avalanche multiplication.

Transport plays an important role for the runaway suppression only at very early
times, before closed flux surfaces have formed (which occurs when the current
reaches $I_{\rm p}\approx\SI{100}{kA}$ at about $t\approx\SI{40}{ms}$). Beyond
this point, confinement rapidly improves and enables net runaway production.
As shown in figure~\ref{fig:REabc_confinementtime}, the losses of runaways due
to drifts remain negligible throughout the simulation, even as drift losses
become the dominant loss mechanism.

In column C of figure~\ref{fig:REabc} we show the evolution of the same case as
in column B, but with all runaway generation disabled (i.e.\ setting
$\partial\nre/\partial t=0$ in equation~\eqref{eq:dnredt}). In this case,
burn-through succeeds and the plasma current and temperature are effectively
ramped up. The large ratio $E/\ED$ is however a major warning sign, as it
quickly rises beyond $E/\ED=\SI{5}{\percent}$ where significant runaway generation would
generally be expected. This illustrates the need to not just ensure that
burn-through is reached for successful start-up, but also to verify that
$E/\ED$ remains sufficiently low for negligible runaways to be produced. Note
that $E/\ED$ should be a better parameter to track in start-up simulations,
rather than the threshold parameter $E/\Ec$, due to the relative importance of
the Dreicer mechanism in these scenarios, which is exponentially sensitive to
$E/\ED$.

\begin{figure}
	\centering
	\begin{overpic}[width=0.67\textwidth]{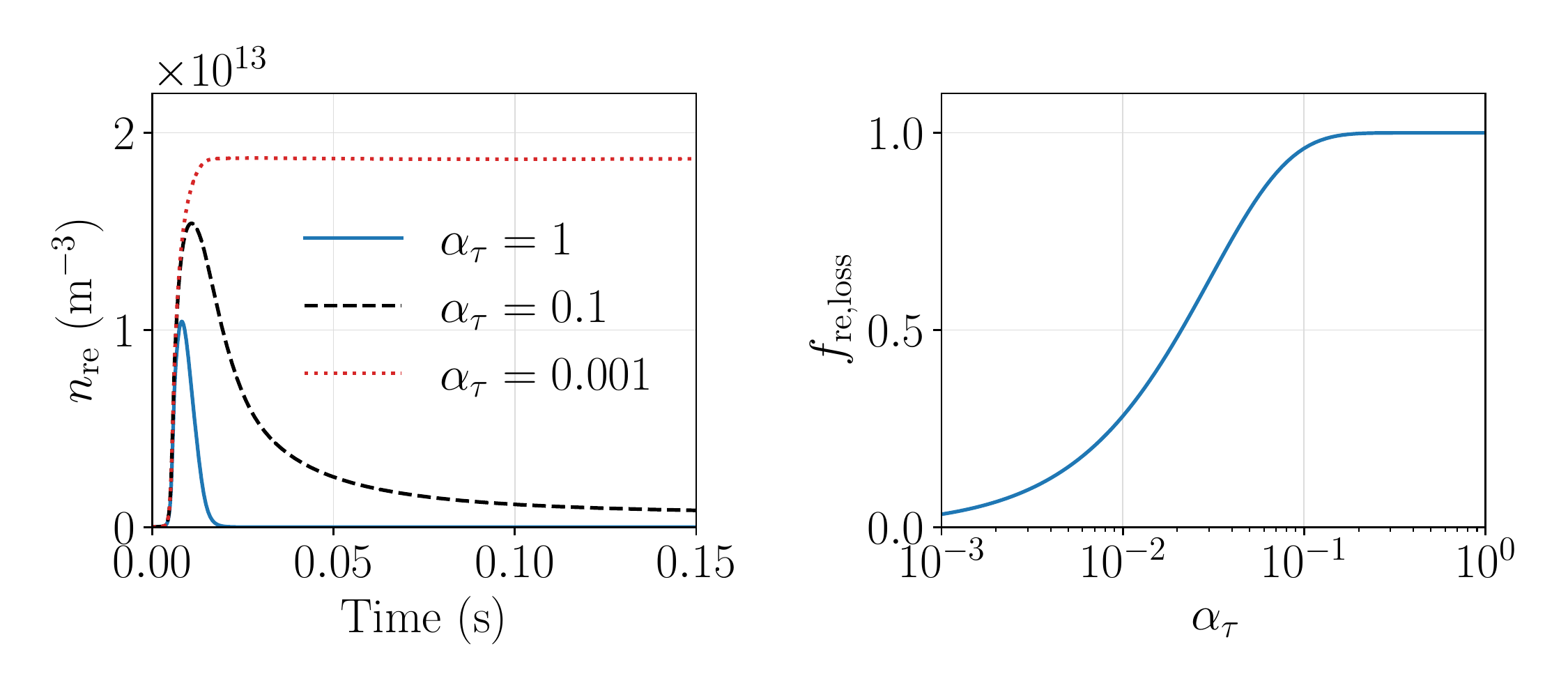}
		\put(11,35){(a)}
		\put(61,35){(b)}
	\end{overpic}
			\caption{
		(a) Time evolution of the runaway electron density $\nre$ for different
		values of the runaway transport scale factor $\fu$.
		(b) Runaway electron loss fraction $f_{\rm re,loss}$,
		defined according to equation~\eqref{eq:floss}, as a function of the
		scale factor $\fu$.
	}
	\label{fig:taurescan}
\end{figure}

One of the most crucial uncertain parameters of the simulation is $\taure$, the
runaway electron confinement time. In section~\ref{sec:remodel} we derived a
heuristic model for the runaway confinement time, and as observed in the
simulations above, it is short enough to completely eliminate the seed runaways
produced in the burn-through phase. The model is however not validated against
experiment, and it could therefore be informative to investigate how sensitive
our results are to variations in this parameter. To this end, we introduce a
scale factor $\fu$ in the transport term of the runaway density
equation~\eqref{eq:dnredt}, such that $\taure\to\fu\taure$. The result of
varying $\fu$ between $10^{-3}$ and 1 is shown in figure~\ref{fig:taurescan},
illustrated by the time evolutions of the runaway electron density $\nre$ for
different $\fu$ and runaway loss fraction
\begin{equation}\label{eq:floss}
	f_{\rm re,loss}=
		\frac{\int_0^{t_{\rm max}}\dd t\left(\gamma_{\rm Dreicer}+\Gamma_{\rm ava}\nre\right)}
		{\int_0^{t_{\rm max}}\dd t\,\nre/\taure},
\end{equation}
where $t_{\rm max}=\SI{0.15}{s}$ was used for these simulations. As previously
observed, with $\fu=1$ (transport exactly according to the heuristic model) all
runaway electrons are lost from the plasma. This remains true for scale factors
down to about $\fu\approx0.2$, where some runaway electrons are able to survive
until the closed flux surfaces form. With $\fu\approx 10^{-3}$, effectively all
runaways generated in the early phase survive the simulation.

\subsection{Runaway electron generation after gas fuelling}\label{sec:fueling}
Since the results of the previous section show that the most crucial parameter
for the generation of runaways is $E/\ED$, we can consider this
parameter to try to understand how runaway electrons could potentially be
prevented. The quantity $E/\ED$ depends on three physical parameters, namely
the electric field strength $E$, the electron temperature $\Te$, and the
electron density $\nel$. Since the goal of the start-up is to reach a target
plasma current and temperature, limiting $\Te$ can only be a temporary measure
and is as such unfeasible for preventing runaways. Limiting $E$ can be effective
in preventing the formation of runaways, but it will also limit the rate at
which the current and temperature can be increased. This leaves the electron
density as the main control parameter. Since $E/\ED\propto\nel^{-1}$, increasing
the electron density (and thereby the electron collision frequency) could
potentially be used to limit the growth of runaway electrons.

As was already illustrated by case A in section~\ref{sec:iterfirst}, a high
prefill pressure can be effective in limiting the runaway electron growth.
However, too high a prefill pressure will also prevent successful burn-through.
Since persistent runaways tend to be generated during later stages of start-up,
after burn-through has occurred, one could imagine a start-up scenario in which
the prefill pressure is kept low to guarantee burn-through, with the density
being subsequently increased to lower, or at least maintain, $E/\ED$ at safe
levels.

\begin{figure}
	\centering
	\begin{overpic}[width=0.71\textwidth]{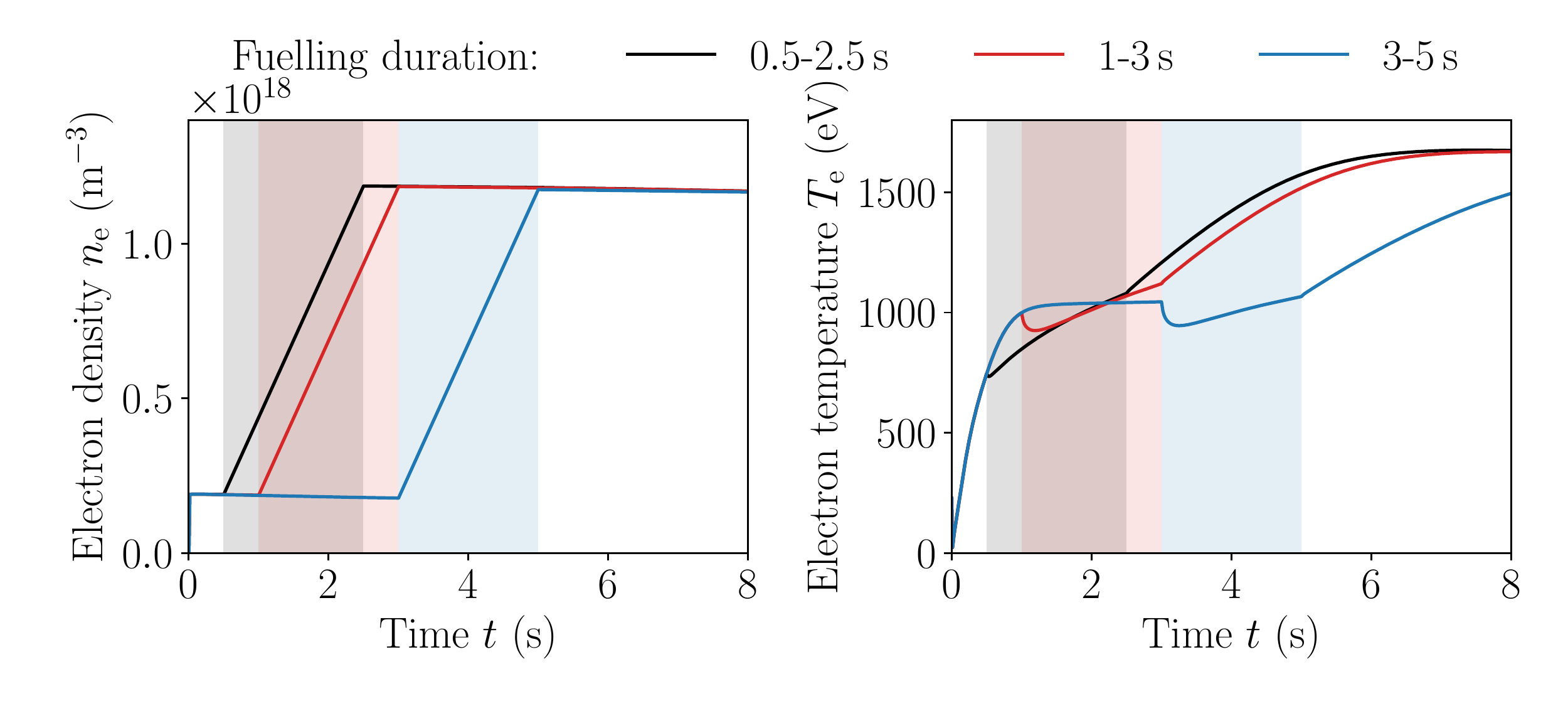}
		\put(14,33.5){(a)}
		\put(62.5,33.5){(b)}
	\end{overpic}
	\caption{
		Evolution of (a) electron density, and (b) electron
		temperature when neutral deuterium is continuously injected for a
		period of two seconds. The initiation of the fuelling is varied between
		the three different cases, and it is started at $\SI{0.5}{s}$ (black),
		$\SI{1.0}{s}$ (red) and $\SI{3.0}{s}$ (blue) respectively. The shaded
		regions correspond to the intervals during which the
		fuelling is active in the different cases.
	}
	\label{fig:gaspuff:plasma}
\end{figure}

In figure~\ref{fig:gaspuff:plasma}, we consider the same scenario as case B
of section~\ref{sec:iterfirst}, but with the neutral
influx~\eqref{eq:neutral:influx} modified to include a source function
$S_{i,{\rm fuel}}^{(0)}$:
\begin{equation}
	\Gamma_{i,{\rm in}}^{(0)} = 
        V_{\rm p}\sum_k\sum_{l\geq 1}
        \frac{Y_k^i n_k^{(l)}}{\tau_k}
		+ \frac{\hat{V}_i^{(0)}}{V_i^{(0)}} S_{i,{\rm fuel}}^{(0)}.
\end{equation}
Here we take the source function for deuterium to be a box function
\begin{equation}
	S_{D,{\rm fuel}}^{(0)}(t) = n_{D,0}\begin{cases}
		1,&\quad t_0\leq t\leq t_0+\Delta t,\\
		0,&\quad \text{otherwise},
	\end{cases}
\end{equation}
with $n_{D,0}=\SI{5e17}{m^{-3}s^{-1}}$, $t_0$ is the activation time and
$\Delta t$ is the duration of the source. As shown in 
figure~\ref{fig:gaspuff:plasma}a, in this study we only vary the onset $t_0$ of
the source (i.e.\ its magnitude and duration are kept constant), and it
results in the same number of injected electrons in all cases. In this section
we consider injections starting at $\SI{0.5}{s}$, $\SI{1}{s}$ and $\SI{3}{s}$
after breakdown, all with a duration of $\SI{2}{s}$.

\begin{figure}
	\centering
	\begin{overpic}[width=\textwidth]{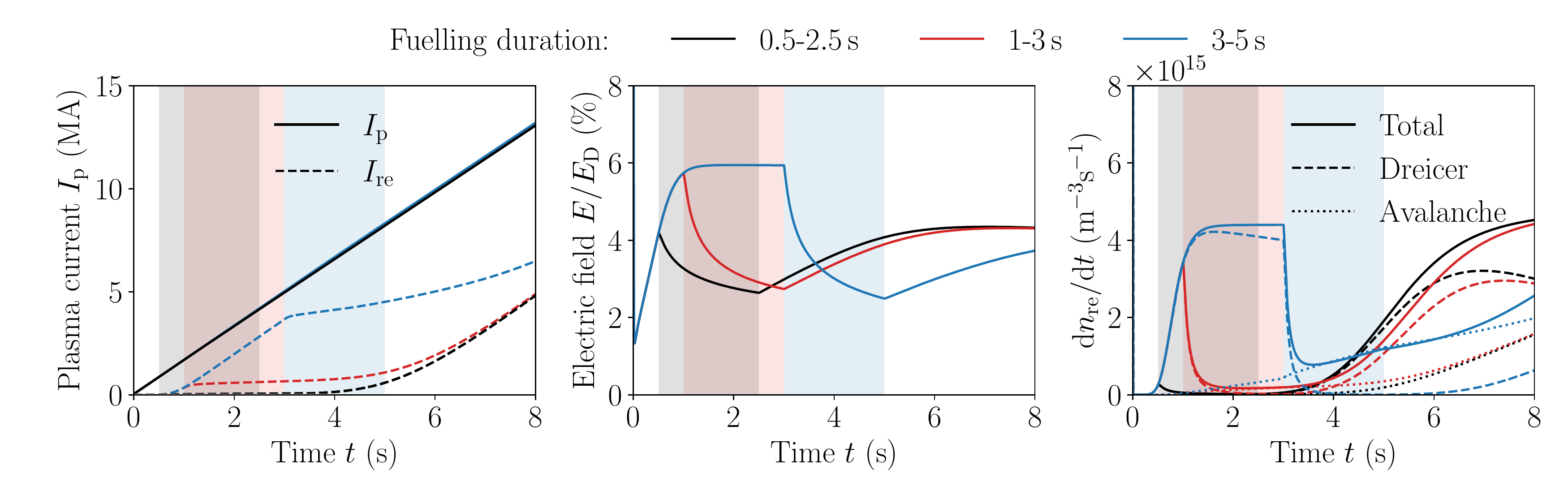}
		\put(10,24){(a)}
		\put(42,24){(b)}
		\put(74,24){(c)}
	\end{overpic}
	\caption{
		Evolution of (a) plasma and runaway current, (b) electric field, and
		(c) runaway generation rate $\dd\nre/\dd t$ in the gas fuelling
		scenarios. By activating the fuelling earlier, the quantity $E/\ED$ can
		be kept down to suppress the Dreicer generation, effectively delaying
		the growth of the runaway electron beam.
	}
	\label{fig:gaspuff:re}
\end{figure}

Figure~\ref{fig:gaspuff:re}a shows the evolution of the total plasma current
$\Ip$ and runaway current $\Ire$. While $\Ip$ remains almost exactly the same
in all three cases, the fraction of the current carried by runaways differs
immensely. This also implies that the ohmic component of the current differs
greatly, which will affect the ability of the electric field to heat the
plasma. In the case with late fuelling (3-$5\,\si{s}$, blue), significant Dreicer
generation occurs which causes runaway electrons to carry $\sim \SI{75}{\percent}$ of the
total current by the time the fuelling is initiated. The parameter $E/\ED$
plateaus near $E/\ED\approx \SI{6}{\percent}$ due to the poor ohmic coupling, allowing the
plasma current to continue to increase while the temperature and ohmic current
remain constant. Once fuelling commences, the Dreicer generation is suppressed as
illustrated in figure~\ref{fig:gaspuff:re}c, but due to the large seed of
runaways already created, and the difficulty of raising the threshold electric
field $\Eceff$ sufficiently high, the runaways can still multiply via the
avalanche mechanism and raise the runaway current further.

When the fuelling source is initiated earlier, at $\SI{1}{s}$ after breakdown,
the effect on the runaway current is significant. Again, the fuelling is
successful in suppressing the Dreicer generation, but in contrast to the delayed
fuelling case, in this case the runaway seed is much smaller. As a result,
avalanche multiplication is much slower and gives the ohmic current time to
rise without runaway electrons overtaking the current. Eventually, after the
gas fuelling is finished and the temperature has risen further, $E/\ED$ recovers
somewhat and again allows Dreicer generation to produce more runaways.

By initiating the fuelling source even earlier, just $\SI{0.5}{s}$ after
breakdown, the effect on the Dreicer generation becomes even more pronounced.
In this case, almost all Dreicer runaways are suppressed, preventing them
from multiplying. Also here, $E/\ED$ eventually recovers and allows the Dreicer
generation to commence again, but only after the ohmic plasma current and
temperature have been raised significantly. Adding to this the fact that, in
reality, the loop voltage will have to decrease significantly after burn-through
\citep{de_Vries_2019}, $E/\ED$ (and by extension the runaway generation) will be
much lower at later times than in our simulations.

From the above analysis we can conclude that gas fuelling may be effective in
preventing significant runaway generation. Our simulations suggest that the
timing of the fuelling is crucial, so that a higher density is reached as soon
as possible after burn-through in order to reduce the parameter $E/\ED$ to
which the Dreicer generation mechanism is exponentially sensitive. If the
density can be raised sufficiently high early on during the start-up, the
potential avalanching of runaways can be greatly reduced and bring the plasma
into an essentially runaway-free operating state.

    \section{Discussion and conclusions}
    \label{sec:conclusions}Runaway electrons produced during plasma initiation can have a significant
effect on the evolution of plasma parameters, in particular in future tokamaks,
where due to restrictions on the allowed toroidal electric field, the prefill
pressure needs to be kept low in order for burn-through to be achievable. The
low prefill pressure, and subsequent plasma density, leads to low values for
the runaway electron threshold electric field $\Ec$, which makes significant
runaway generation much more likely. Since most plasma parameters vary greatly
during tokamak start-up, it is crucial to evolve the plasma parameters
self-consistently in simulations of the start-up. In this study, we have
therefore developed the new simulation tool \STREAM\ to self-consistently couple
models for the background plasma to models for the runaway electrons. The
background plasma models have been benchmarked to the results of \DYON\
presented in \citep{Kim2020}, which were in turn compared to simulations of
the same scenarios using \BKD\ and \SCENPLINT, and show good agreement.

In section~\ref{sec:iterfirst} we considered an ITER first plasma-like scenario
and varied the prefill pressure, as well as whether or not runaway electron
generation was accounted for in the model. We found that a low prefill pressure
would lead to significant runaway generation, which would also impact the ohmic
coupling between the electric field and the plasma, thus limiting the amount of
heat which could be provided to the plasma in the presence of runaways. The
clearest sign of whether significant runaway generation would occur in our 
simulations was when $E/\ED$, the electric field normalized to the Dreicer
field, reached values of more than $\sim \SI{3}{\percent}$ during a significant
amount of time {\em after} successful burn-through. Prior to burn-through,
$E/\ED$ usually takes extremely high values, but any runaway electrons which are
generated during this phase are almost immediately lost to the wall due to the
open field line configuration. Considering that significant runaway electron
generation can impact the plasma properties during start-up, monitoring the
evolution of $E/\ED$ and ensuring that it is kept sufficiently low could be a
straightforward way to verify the applicability of other existing burn-through
models.

The confinement time for runaway electrons is a highly uncertain parameter in
our simulations, and they suggest that the result that all runaway electrons are
lost is robust to within a factor of $\sim 10$ in the confinement time. This
assumes that the generation is dominated by relativistic electrons, something
which is not necessarily the case in an experimental setting. It is possible
that electrons only attain moderate superthermal energies, likely leading to
improved confinement of these electrons which are prone to continue accelerating
and eventually turning into relativistic runaway electrons.

We also studied the effect of gas fuelling on the generation of runaway electrons
and found that injection of deuterium after burn-through can be effective in
suppressing much of the runaway generation. The large values of $E/\ED$
obtained early on in low prefill scenarios gives rise to significant Dreicer
generation. This provides a large seed of runaway electrons which will avalanche
even after $E/\ED$ has been reduced by the increased density, thus allowing
a significant runaway current to build up. By fuelling the plasma early after
burn-through, the Dreicer generation can be suppressed, leaving little or no
runaway electrons available to avalanche when $E/\ED$ eventually recovers.
By increasing the density even further than we have done in this paper it
might also be possible to raise the runaway threshold $\Ec$ above the applied
electric field $E$, thus also suppressing the avalanche multiplication.
However, note that if fuelling is increased too much or too early, plasma
burn-through will fail for the same reason as high prefill gas pressure cases.
This calls for careful optimization of the prefill gas and fuelling to avoid the
risk of both runaways and failed burn-through.

It is important to note that our studies are qualitative rather than
quantitative, and that we have not considered the response of the tokamak
control system in these simulations. Since the plasma resistivity decreases with
temperature, the electric field should eventually be limited to maintain a
constant plasma current when the target temperature is reached. When this
happens, it should limit the avalanche generation mechanism, stop further
runaway growth and instead reduce the number of runaways, if any, via
collisions.

The focus in section~\ref{sec:fueling} has been on preventing the generation of
runaway electrons. A natural follow-up question to ask would be if it is
possible to suppress already existing runaways. In our model, the only
reasonable means for reducing the number of runaway electrons is to reduce the
threshold parameter $E/\Eceff$ below unity to allow the runaways to avalanche
``in reverse'' and collisionally dissipate their energy to the bulk electrons.
This might be feasible, and a natural result of successful start-up, if the
fraction of runaway current is low. If, however, the runaway current is
significant, this process would be difficult to reconcile with maintaining the
plasma at a specified current and temperature.

In this study we have only considered the effect of relativistic
electrons.  During start-up it is also possible for the electron
distribution function to be significantly distorted from thermal
equilibrium due to the large values of $E/\ED$, and for a large
fraction of superthermal (albeit not relativistic) electrons to
form. Such electrons will also affect the heating properties and
general evolution of the plasma, but are not captured by the model
presented here. An analysis of these electrons would require the
solution of the Fokker--Planck equation with non-linear collision
coefficients, since linearizations of the collision operator typically
assume that most electrons are in thermal equilibrium. While \DREAM,
and thus by extension \STREAM, supports the solution of a
Fokker--Planck equation simultaneously with the evolution of the
(fluid) background plasma, all collision operators available in the
code are linearized. A kinetic treatment of the runaway problem during
start-up could also allow us to study the effect of runaways on the
ionization of atoms, as well as the effect of Electron Cyclotron
Heating (ECH)---the latter which is anticipated to be used in ITER to
assist startup---for the generation of fast electrons. This could be
done by, for example, coupling \STREAM\ to the bounce-averaged
Fokker-Planck code \LUKE\ \citep{Decker2004} which also solves for
wave-particle interactions. We therefore leave for future studies a
more detailed analysis of the momentum-space dynamics of fast electrons
during start-up.

    \section*{Acknowledgements}
    The authors are grateful to H.~T.~Kim, E.~Nardon, S.~Newton, I.~Pusztai
	and J.~Decker for fruitful discussions.

	\section*{Funding}
        This work has been carried out within the framework of the EUROfusion
    Consortium, funded by the European Union via the Euratom Research and
    Training Programme (Grant Agreement No 101052200 — EUROfusion). Views and
    opinions expressed are however those of the author(s) only and do not
    necessarily reflect those of the European Union or the European Commission.
    Neither the European Union nor the European Commission can be held
    responsible for them. This work was supported by the Swedish Research
	Council (Dnr. 2018-03911). This work was supported in part by the Swiss 
	National Science Foundation.

	\section*{Declaration of interests}
	The authors report no conflict of interest.

	\section*{Data availability statement}
	The source code for \STREAM\ as well as simulation scripts for all
	simulations of this study are openly available in
	\url{https://github.com/chalmersplasmatheory/STREAM}.

    \bibliographystyle{jpp}
    \bibliography{ref}
    
\end{document}